Structural study in Highly Compressed BiFeO$_3$ Epitaxial Thin Films on YAlO$_3$

Heng-Jui Liu,[1,2] Hsiang-Jung Chen,[3] Wen-I Liang,[3] Chen-Wei Liang,[3] Hsin-Yi Lee,[2] Su-Jien Lin,[1] Ying-Hao Chu[3,a)]

[1]Department of Materials Science and Engineering, National Tsing Hua University, Hsinchu 30013, Taiwan

[2]National Synchrotron Radiation Research Center, Hsinchu 30076, Taiwan

[1]Department of Materials Science and Engineering, National Chiao Tung University, Hsinchu 30010, Taiwan

Abstract

We report a study on the thermodynamic stability and structure analysis of the epitaxial BiFeO$_3$ (BFO) thin films grown on YAlO$_3$ (YAO) substrate. First we observe a phase transition of M$_C$–M$_A$–T occurs in thin sample (<60 nm) with an utter tetragonal-like phase (denoted as M$_{II}$ here) with a large *c/a* ratio (~1.23). Specifically, M$_{II}$ phase transition process refers to the structural evolution from a monoclinic M$_C$ structure at room temperature to a monoclinic M$_A$ at higher temperature (150$^o$C) and eventually to a presence of nearly tetragonal structure above 275$^o$C. This phase transition is further confirmed by the piezoforce microscopy measurement, which shows the rotation of polarization axis during the phase transition. A systematic study on structural evolution with thickness to elucidate the impact of strain state is performed. We note that the YAO substrate can serve as a felicitous base for growing T-like BFO because this phase stably exists in very thick film. Thick BFO films grown on YAO substrate exhibit a typical "morphotropic-phase-boundary"-like feature with coexisting multiple phases (M$_{II}$, M$_I$, and R) and a periodic stripe-like topography. A discrepancy of arrayed stripe morphology in different direction on YAO substrate due to the anisotropic strain suggests a possibility to tune the MPB-like region. Our study provides more insights to understand the strain mediated phase co-existence in multiferroic BFO system.

I. Introduction

Multiferroics have been widely studied due to their fascinating physical properties, such as the coupling between electric and magnetic orders[1,2] and conduction in domain walls[3], offering an opportunity for novel devices. Among those multiferroics, BiFeO$_3$ (BFO) is the most promising one because of its high curie temperature of ferroelectric order ($T_c$=1103K) and antiferromagnetic order ($T_N$=643K).[4-6] Recently, it has been shown a strain-driven conceptual "morphotropic phase boundary"[7] (MPB) with superior spontaneous polarization in the epitaxial thin film BFO under highly compressive strain (~-4%) [8,9]. Generally, MPB describes a phase transition from tetragonal (T), monoclinic (M) to rhombohedral (R) symmetries induced by compositional change, which is usually observed in the lead-based ferroelectrics. These compounds captured significant attention due to the strongly enhanced ferro/piezoelectricity.[10-12] However, Pb-based compounds are not environmentally friendly, strain-driven MPB suggests a new avenue to design new green ferro-/piezoelectrics.

In the previous studies[7,15-17], the MPB-like region in the strained BFO thin films is composed at least two phases, which are usually labeled as $M_{II}$ (or tetragonal-like) and R (rhombohedral-like) phases.[15] High-resolution x-ray diffraction techniques disclosed the monoclinic nature of $M_{II}$ phase and defined precisely that the R phase is similar to rhombohedral structure presented in bulk BFO yet the degree of distortion presents larger. In addition, an extra phase with monoclinic or triclinic structure is found to accommodate the lattice difference between $M_{II}$ and R BFO, which is labeled as $M_I$.[15,16] These phase stacking sequence of MPB-like structure could also be connected to the sawtooth-like morphology observed from atomic force microscopy (AFM) and transmission electron microscopy (TEM).[7,15,17] On the other hand, plenty of studies exuberantly focused on the understanding of those bridge phases, and they further inferred that the $M_{II}$ phase in fact belongs to the $M_C$ structure (*Cm* or *Pm* symmetry) with a shear angle along [100] direction. When the compressive strain is reduced by environmental change such as thickness or substrates, $M_{II}$ phase transforms into R-like phase, which is later confirmed to be $M_A$ or distorted rhombohedral structure (*Cc* or *R3c* symmetry) with a shear angle rotation from [100] to [110] direction, a process that induces a spontaneous polarization rotation correspondingly.[17-21] This structural transition has been considered as the main contributor to the piezoelectric anomaly in the BFO/LAO system.

While detailed information of phase transition in BFO has been deeply discussed in the system of BFO/LAO, there were not many works about highly strained BFO thin films on YAlO$_3$(YAO) substrates (a=c=3.70Å b=3.67Å, *β*=88.47° in pseudocubic setting), which the lattice should be more suitable for the predicted tetragonal structure of BFO (*P4mm* symmetry with a=3.66Å, c=4.67Å)[22,23]. In this study, we replace the LAO substrates with YAO substrates to scrupulously inspect the variations of MPBs between these two substrates. We observe a $M_C$–$M_A$–T phase transition via temperature-dependent experiments, which is distinct to the case of BFO on LAO substrates [25,26]. Our study also builds a universal view of polarization rotation undergoing the $M_C$–$M_A$ phase transition. In latter part of study, we turn our focus onto the thicker film with MPB-like region and elucidate the existence of in-plane strain anisotropy in the BFO/YAO system that provides a new mean to control the MPB-like region. With the aim of understanding the inherent influence of in-plane strain anisotropy offered by YAO substrates, we study and compare the entire evolution of BFO lattice parameters with thickness variation on LAO and YAO substrates.

II. Experimental process

Epitaxial BFO thin films were prepared by reflection high-energy electron diffraction (RHEED)-assisted pulsed laser deposition from a BFO target with 10% excess bismuth. Growth was carried out at 700°C at oxygen pressure of 100 mTorr on single-crystal YAO (110) substrates in orthorhombic index. The thickness of the BFO films was controlled by a combination of RHEED monitoring and deposition time. The films were then cooled in oxygen pressures of approximately 760 Torr. The topography of all samples and their polarization domain structures were studied using atomic force microscopy (AFM, Veeco Escope) and piezoelectric force microscopy (PFM). Tips used for PFM imaging were Ti-Pt coated cantilevers with an elastic constant of 4.5 N/m and a resonance frequency of 120~190 kHz. When performing the PFM measurements, the scanning speed was set at 5 μm/s, the ac excitation frequency was 10.5 kHz, and the ac amplitude was 7 Vpp. The PFM images have been recorded with the tip cantilever scanning along [100] and [010] direction. Structural details of the samples were collected by synchrotron-based X-ray diffraction (XRD) techniques at beamline BL-13A at the National Synchrotron Radiation Research Center (NSRRC) in Hsinchu, Taiwan. The incident beam was monochromated at 12KeV with Si (111) double crystal mirror. Two sets of slits were placed before samples to get the beam size about 0.4mm × 0.8mm and the other two were placed after the sample (or before scintillation counter) to decrease background noises. The x-ray reciprocal space map was measured step by step and plotted in the reciprocal lattice unit that is normalized to YAO substrate in pseudocubic settings (1 r.l.u.=$2\pi/a_{YAO, pc}$).

III. Results and discussion

For understanding the structural evolution of the BFO thin films grown on YAO substrates, two relevant phases of BFO reported before should be mentioned first. For instance, the BFO thin films grown on the substrates with larger lattice mismatch (>4%, related to the lattice of bulk BFO) such as LaAlSrO$_4$, LAO, YAO prefer to form the monoclinic M$_C$ structure illustrated in Fig.1 (a). Those grown on the substrates with smaller lattice mismatch such as SrTiO$_3$ (STO) belong to monoclinic M$_A$ structure (Fig. 1(b)). For M$_C$, the monoclinic distortion is along [100] with respect to pseudocubic structure; therefore, the corresponding domain structures would result in three-fold splitting and double splitting in the reciprocal space mappings (RSMs) of (H00) and (HH0) planes. By contrast, M$_A$ has a monoclinic distortion along [110] in the pseudocubic index, which results in the opposite diffraction patterns in RSM analysis. In general, M$_A$ is also indicated as the R-like phase due to that the unit cell volume is close to the bulk BFO.

We first study the structural evolution of 18 nm BFO thin film on YAO as a function of elevated temperature. The θ−2θ x-ray diffraction scans along [001]$_{pc,YAO}$ direction shown in Fig. 2(a) reveals the existence of M$_{II}$ phase diffraction peaks from room temperature to 400°C. The calculated c-axis lattice parameter of M$_{II}$ phase displays a temperature dependent behavior, which is distinct from that of YAO substrates (Fig. 2(b)). YAO substrate exhibits nearly linear thermal expansion along its normal direction [001]$_{pc}$, whereas the c-axis parameter of BFO thin film varies: it is elongated and then stabilized in the temperature range of 100~250°C, however, it decreases again when temperature mounts higher. This non-linear variation of c-axis lattice parameter of BFO implies that a possible ferroelectric or antiferromagnetic transition occurs due to structural variation[27,28]. In order to study this transition, detailed structure of M$_{II}$ phase is shown by measuring asymmetry RSMs exhibited in Fig. 2(c)-(h). These RSMs (Fig. 2(c) and 2(d)) demonstrate the typical diffraction patterns at room temperature: Three-fold splitting and two-fold splitting are observed respectively around (103) and (113) reflections, recognized as M$_C$ in Fig.1 (a). The lattice parameters of this M$_C$ structure obtained from RSMs are a=3.79Å, b=3.75Å, c=4.63Å and β$_{MC}$=88.83°. We further performed the same measurements at 150°C. One can find that the RSMs of the same sample at 150°C (Fig. 2(e) and 2(f)) display reverse patterns, which present a typical M$_A$ structure feature (Fig.1 (b)): Two-fold splitting and three-fold splitting corresponding to the (103) and (113) reflections. In order to compare with M$_C$, the a-axis and b-axis of M$_A$ are redefined into the M$_C$ coordinate for clarity and the lattice parameters are calculated as a=3.768Å, b=3.761Å, c=4.645Å and β$_{MA}$=87.1°. This provides a direct evidence of phase transition. Besides, this M$_A$ phase sustains high c/a ratio (~1.23), which is different from the previous M$_A$ cases such as BFO thin films on STO substrate (c/a~1.04).[17-21,24] Similar phenomena are also observed in the case of LAO substrates.[25-27] First principle prediction proposed by Íñiguez *et al.* has indicated highly compressive strained BFO thin films show two sets of symmetries for energetically favorable phases.[29] One is the *Pm*, or *Cm* symmetry, where the shear direction is along [100],[18,29] and the other is the *Cc* symmetry, where the shear direction is along [110].[30] Both have a large c/a value and only a little difference in energy, which means these two sets of symmetries for the $M_{II}$ phase can be stably existed on these substrates. Hence, we suggest that the $M_C$ phase would change its symmetry from *Pm* or *Cm* at low temperature to $C_C$ (the $M_A$ phase) at high temperature. If we further increase the temperature to 275°C, the structure seems to become more tetragonal as shown in Fig. 2(g) and (h): there is nearly one diffraction spot, which represents the withdrawal of tilting angle, in the (103) and (113) reflections. Overviewing the heating process, it is apparent to conclude that $M_{II}$ exhibits $M_C$–$M_A$–T phase transition through the study of RSMs.

To verify the $M_C$–$M_A$–T phase transition, PFM measurement was carried out. In general, $M_C$ and $M_A$ have different in-plane polarization directions due to their different shear directions. Due to the self-poling effect (asymmetric electrostatic boundary conditions), the complicated eight polarization domain states can be limited to only four components, as shown by the schematics in Fig. 3(a) and 3(b).[19,31] For $M_C$ phase, the scanning cantilever along with [100] should result in three levels of piezoresponse signals: the two opposite polarizations perpendicular to the cantilever lead to the darkest (red arrow) and brightest contrast (blue arrow); the other two opposite polarizations parallel to the cantilever almost present the same contrasts in the medium color level (green arrow). For $M_A$ phase, only two levels of piezoresponse signals can be expected because for the four variant polarizations, only the polarizations project onto the perpendicular components can be detected when the cantilever scans along in [100] direction.

At room temperature, the surface topography shows a smooth with a little step bunching morphology (Fig. 3(c)). Its in-plane PFM image shows stripe-like feature when scanning along [100] direction (Fig. 3(d)). The stripes are formed to minimize the electrostatic and elastic energy of the ferroelectric domains, hence the domain walls would intersect the {001} plane in the <110> direction to form four disparate domain states, such as the colored indicators in Fig. 3(c).[17,19,32] The PFM results are consistent of Bokov's prediction for the $M_C$ phase.[31] Repeating the AFM and PFM measurements at 150°C (Fig. 3(d)), we find a revolute domain structure occurs while the topography remains the same. A puddle-like feature with dark and bright contrasts and the domain walls that intersect the {001} plane in the <110> indicate a typical $M_A$ feature. The fact that the in-plane polarization direction changes from [100] to [110] at elevated temperature confirms again the strained BFO/YAO system undergoes a phase transition from $M_C$ to $M_A$, which is consistent with the results in RSMs analysis. However, we do not have clear IP-PFM image for the sample at 275°C, so the newly unveiled phase transition of $M_A$-T still needs to be confirmed.

In order to explore the impact of strain state on BFO thin films imposed by YAO substrates, BFO samples with various thicknesses (18 nm to 300 nm) were characterized by RSMs in normal direction shown in Fig. 4(a)-(d). As thickness increases, multiple phases are coexisted, and when the thickness is larger than 60 nm, the coexistence of "$M_I$", "$M_{II}$", "$M_{II,tilt}$" and "R" pictures the view of MPB-like region in BFO. The notations here, except for the $M_{II}$ structure defined before, three extra notations of $M_I$, $M_{II,tilt}$ and R marked in these RSMs are also adapted from previous studies on BFO/LAO system[15,26]. More specifically, "R" refers to the distorted rhombohedral phase of BFO bulk; "$M_I$" and "$M_{II,tilt}$" are the intermediate phases possessing tilted monoclinic structure to accommodate the large lattice mismatch between "$M_{II}$" and "R".[18] In this work, $M_{II}$ appears in all the samples and the mixed phase region starts to be formed above 60 nm, which exhibits the similar sawtooth-like morphology (Fig. 5(b)-(d)) consisting of $M_I$ and $M_{II,tilt}$ phases that happened in BFO/LAO system.[15,26] We further study these BFO phases via the characterizations of rocking curves for each phase. First, we selected the position of $M_I$ phase at L=0.887 (c=4.18Å) as shown in Fig. 4(e). From the result of the rocking curve, only a broader shoulder around ΔH=0 is observed in the thin film with 18 nm. As thickness keeps increasing to 60 nm, two extra peaks are located at ΔH≈±0.05, indicating that $M_I$ phase has a tilting angle about ±3º from the surface normal direction. In addition, we also scrutinized the rocking curves of $M_{II}$ phase shown in Fig.4(f). The $M_{II}$ phase begins to split into three peaks when the thickness reaches above 300nm: one is for normal direction and two outer peaks are $M_{II,tilt}$ phase. It can be speculated that the $M_I$ is the main intermediate phase to balance the lattice mismatches between $M_{II}$ and R, while the R phase begins developing. As the proportion of R raises with increasing thickness, only $M_I$ phase is not enough to compensate increasing height difference from $M_{II}$ to R; hence, some part of $M_{II}$ is dragged by adjacent $M_I$ to form a tilting structure ($M_{II,tilt}$), which results in a sawtooth-like feature.

In Fig. 5(a)-(d), we examined the morphology evolution via scanning probe microscope. When BFO is thin (~18 nm), the topography of pure $M_{II}$ phase is very flat shown in Fig. 5(a). When the film is thick enough, the arrays of the striped regions consisting of periodic arranged stripe-like structure emerges. These stripe-like structures, or MPB-like regions, have been proven to the presence of intimately mixed $M_{II}$, R, $M_I$, and $M_{II,tilt}$ phases. Moreover, the AFM images of these stripes on YAO substrates show a different feature and strain release process compared to those on LAO substrates, especially in the thickest film (300 nm). In the case of BFO/LAO system (Fig. 5(e)), we could barely find the difference between the stripe-like feature along [010] and [100] direction of LAO substrate. The stripe-like feature on the YAO substrate (Fig. 5(d)), interestingly, exposes an anisotropy behavior. For clarity, we focus on the thickest sample to speculate the discrepancies from two orthogonal direction of YAO substrate (namely, one is $[010]_{pc}$ and another is $[100]_{pc}$). Along $[010]_{pc}$ direction, the stripes arrangement has a commodious spacing ( approximately 6.5 stripes per unit length in average) with more deeper valley structures (about 2.4 nm, statistically speaking). On the contrary, those stripes along $[100]_{pc}$ possess a very dense (almost 10 stripes per unit length), shallow (about 1.6nm in average) morphology. Through careful measurements on the these features shown in Fig. 5(f), the stripes along $[010]_{pc}$ direction show that one side inclines about $1.8^o$ from film surface and the other side inclines about $2.9^o$, corresponding to the $M_{II,tilt}$ and $M_I$ phases from XRD rocking curves, respectively. On the contrary, the stripes along $[100]_{pc}$ direction show that one side inclines about $1.91^o$ and the other side inclines about $2.61^o$. This phenomenon is confirmed by the rocking curves of thickest film presented in Fig. 4(e) and (f). One can find that the peaks of $M_I$ phase along $[100]_{pc}$ direction are located at outer position than those along $[010]_{pc}$ direction, implying the $M_I$ phase along $[100]_{pc}$ tilts more than $[010]_{pc}$ direction. The opposite situation can be observed in the rocking curves of $M_{II}$ phase. The tilted angles of $M_I$ phase are then extracted from rocking curves, which are $2.96^o$ and $2.7^o$ along $[100]_{pc}$ and $[010]_{pc}$, respectively. And those of $M_{II,tilt}$ are $1.27^o$ and $1.46^o$ along $[100]_{pc}$ and $[010]_{pc}$, respectively. If we gather further statistics of the valley volume per unit length, one could derive that the volume ratio of $[010]_{pc}$ and $[100]_{pc}$ is about 1.33, which also implies that the strain energy released in $[010]_{pc}$ direction is larger than that of $[100]_{pc}$ direction. From the understanding of $M_{II}$ phase and YAO pseudocubic lattice setting discussed above, both in-plane lattices of $M_{II}$ are larger than those of YAO, and the domains along $[010]_{pc}$ direction would suffer larger stress than those along $[100]_{pc}$, representing that more elastic energy should be released along $[010]_{pc}$ to sustain the completeness of $M_{II}$ structure everywhere. Therefore, the information mentioned above serves as a guide for the fact that the arrayed stripe anisotropy should be directly related to the lattice difference of the a-axis and b-axis of the YAO substrate.

The structural development of dominant phase $M_{II}$ is crucial to understand the the progress of the strain state because formation of MPB-like feature functions as the outlet of large compressive strain. In this passage, we discuss about the entire trend of $M_{II}$ behavior. The a/b ratio of the $M_{II}$ phase is calculated to be about 1.01 shown in Fig. 6(a), illustrating the effect of biaxial compressive strain. From Fig. 6(b), one also can perceive that the shear angle β decreases with increasing thickness and the c/a ratio keeps almost the same (about 1.235 above 60nm). These observations suggest that the relaxation of the compressive strain goes through two ways on YAO substrate, which is relaxed by the recovery of lattice constants first and then by the more shear distortion of lattice structure. Secondly, the in-plane lattice parameters obtained from symmetry and asymmetry reflections are listed in Table. I. The reduced in-plane lattice (a- and b-axis) and elongated out-of-plane lattice (c-axis) accompanying with increasing thickness elucidate that the BFO films grown on YAO substrate should sustain the compressive stress. Besides, we compare the similar thickness experiment on LAO substrates performed by Zeches *et al*. in the lower part of Table. I. The obvious shrinkage of cell volume on LAO substrates indicates that $M_{II}$ would fully relax when thickness exceeds 80nm and almost converts into R phase above 260nm, while on YAO substrate, the cell volume sustains nearly constant in samples with thickness larger than 300 nm. Therefore, the critical thickness of the $M_{II}$ phase on YAO substrate is much larger, and meanwhile, the YAO substrate offers a better environment for $M_{II}$ phase growth.

IV. Conclusion

In conclusion, we have demonstrated that highly compressively strained BFO thin films can be successfully grown on YAO substrate with the presence of $M_{II}$ (T-like) phase. The structural details from symmetry and asymmetry XRD analysis manifest that the $M_{II}$ phase has a monoclinic structure $M_C$ at room temperature, and then goes through a new kind of phase transition $M_C$-$M_A$-T when the temperature is elevated. The PFM results confirm the $M_{II}$ phase indeed transforms from $M_C$ to $M_A$ through the in-plane polarization rotation from [100] to [110]. BFO thin films with various thicknesses are investigated to draw out the minute picture. With increasing film thickness, BFO thin films gradually exhibit the typical mixed phase feature. The surface morphology and XRD studies reveal that stripes consisting of sawtooth-like periodic arrays have nearly the same phases ($M_I$ and R) with respect to those on LAO substrates. And for the merits of YAO substrate, the dominated phase $M_{II}$ shows almost the same unit cell volume with thickness exceeding 300nm. This suggests YAO is a suitable environment for T-like BFO growth. Nevertheless, in our work, we have manipulated a substrate-induced anisotropy to serve as a method to tune the morphology of stripe arrays: denser and shallower stripes prefer to arrange along YAO $[100]_{pc}$, and broader and deeper stripes prefer to arrange along YAO $[100]_{pc}$. We can associate the phenomena with the little difference presented in YAO in-plane lattice parameters.

Acknowledgment

Financial support of the National Science Council through project NSC 100-2119-M-009-003 is gratefully acknowledged by the authors.

Table 1. Lattice parameters of monoclinic BFO phase grown on YAO and LAO substrates with different thickness

| Substrate | thickness | a (Å) | b (Å) | c (Å) | $\beta$ (°) | V(Å³) |
|---|---|---|---|---|---|---|
| YAO | 18nm | 3.78(9) | 3.75(1) | 4.63(2) | 88.83 | 65.81(9) |
| | 66nm | 3.76(5) | 3.73(2) | 4.65(1) | 88.54 | 65.33(0) |
| | 180nm | 3.76(6) | 3.72(7) | 4.65(2) | 88.24 | 65.26(4) |
| | 300nm | 3.76(5) | 3.72(4) | 4.65(7) | 88 | 65.25(5) |
| LAO* | 17nm | 3.83(8) | 3.77(2) | 4.62(7) | 88.67 | 66.69(1) |
| | 53nm | 3.84(4) | 3.75(3) | 4.64(4) | 88.51 | 66.79(3) |
| | 89nm | 3.66(7) | 3.58(6) | 4.69(1) | 88.63 | 61.58(2) |
| | 120nm | 3.66(5) | 3.59(1) | 4.70(1) | 88.79 | 61.85(6) |
| LAO** | 260nm | 3.95(5) | 3.95(5) | 3.97(1) | 89.4 | 61.94(1) |

\* ref. 7 supporting material

\*\* ref. 26

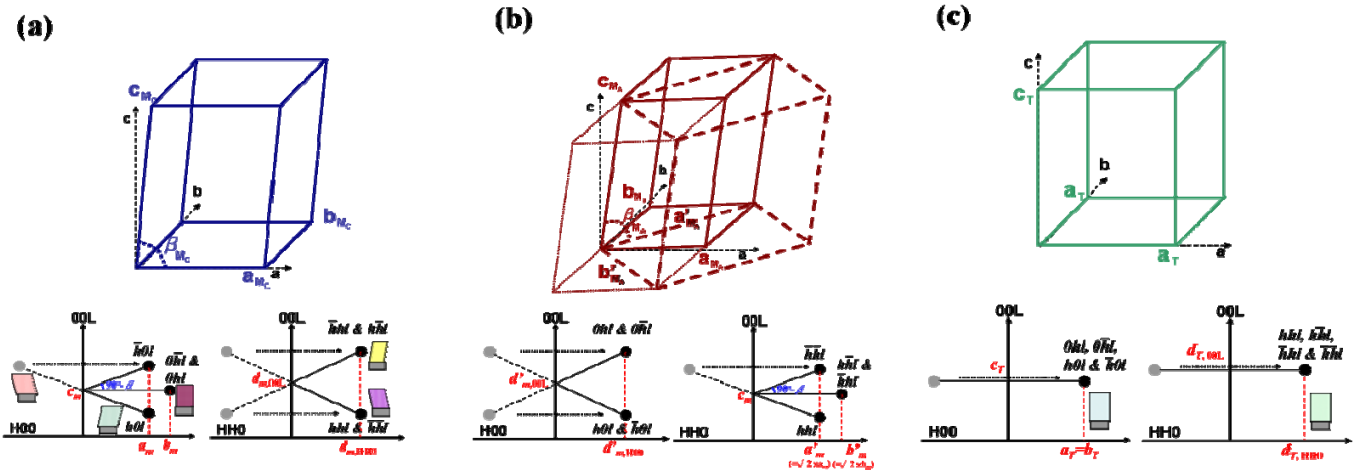

FIG. 1. The schematics of each structure and the corresponding diffraction features:(a) $M_C$, with the shear orientation along [100] direction, (b)$M_A$, with the shear orientation along [110], and (c)T phase, without shear angle. The shear angle will cause the peak splitting due to the four kinds of domains, and its direction will results in the reverse patterns in the (H0L) and (HHL) scattering zones.

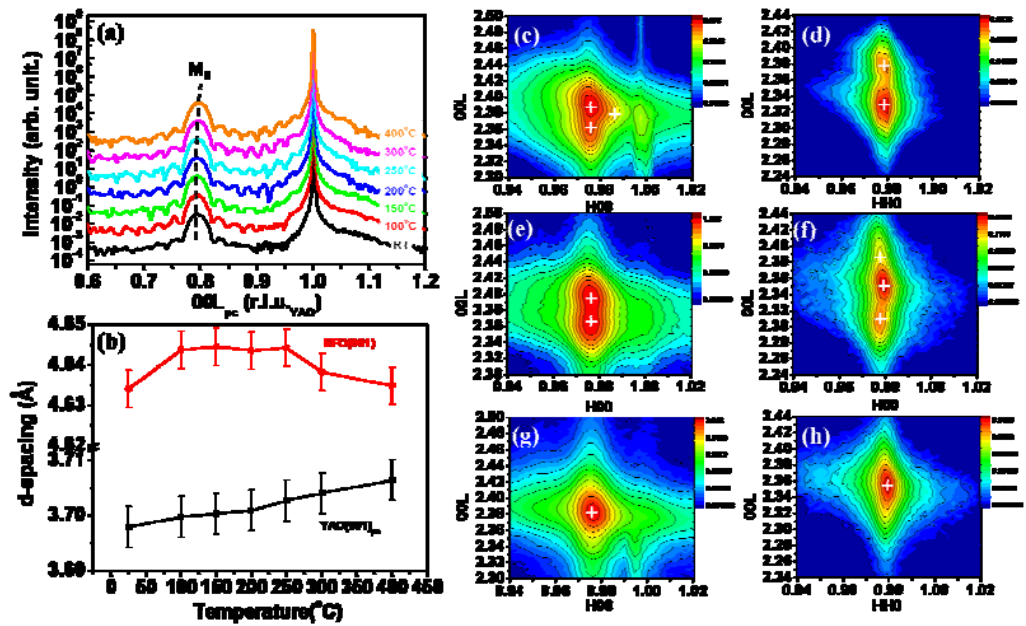

FIG. 2 (a) X-ray normal scan of a BFO thin film with reference to YAO (001)$_{pc}$ peaks at various temperatures. Dashed line is used as a guide to visualize the shifts of the M$_{II}$ peaks. (b) *c*-axis lattice parameters of BFO (001) and YAO (001)$_{pc}$ from RT to 400°C. (c) and (d) show experimental RSMs of (103) and (113) of our BFO thin film at RT. (e) and (f) are the RSMs of (103) and (113) of the BFO thin film at 150°C. (g) and (h) are the RSMs of (103) and (113) of the BFO thin film at 275°C. These RSMs unveiled the phase transition of M$_C$-M$_A$-T.

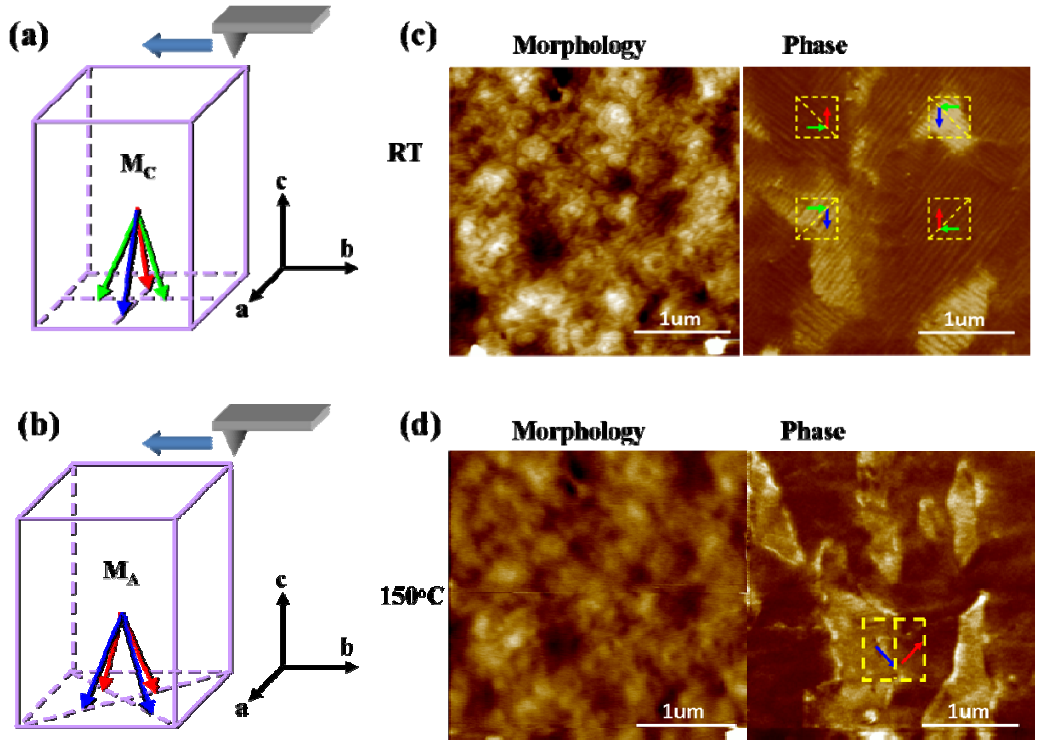

FIG. 3. (a) Schematics of the ferroelectric polarizations in $M_C$, which shows four kinds in-plane polarization variants on the {100} planes. Three contrasts (blue, green, and red) are expected from PFM measurements when the cantilever is aligned to [100]. (b) Schematics of the ferroelectric polarizations in $M_A$, which also have four in-plane polarization variants on {110}. However, when conducting PFM measurements with the cantilever aligned to [100], two contrasts (dark, light) are expected in $M_A$. (c) AFM and PFM phase images of a BFO sample at RT with the cantilever aligned to [100]. (d) AFM and PFM phase images of the same scanning area and direction as shown in (c) but at 150°C. The blue, green, and red arrows in phase images of (c) and (d) indicate the directions of the in-plane polarization variants illustrated in (a) and (b). The combinations of these polarization variants would cause the stripe-like and puddle-like domain features in $M_C$ and $M_A$, respectively.

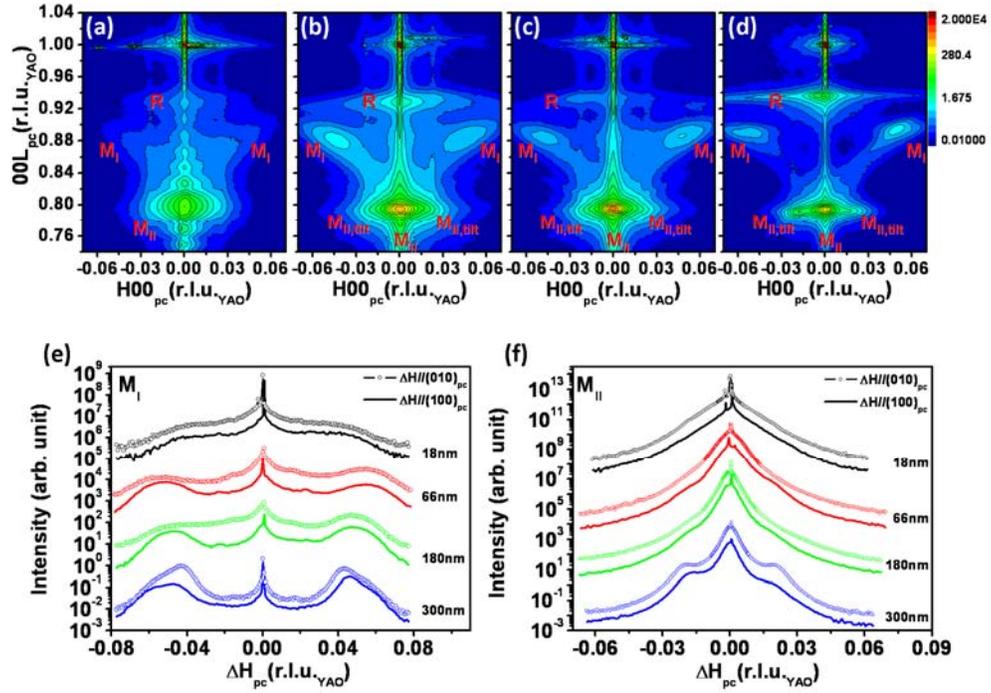

FIG. 4. (a)-(d) X-ray normal reciprocal space maps (RSM) of thin films with various thicknesses. $M_{II}$, $M_{II,tilt}$, $M_I$, and R phase are marked near the position of relevant peaks. (e) and (f) are the rocking curves of $M_I$ and $M_{II}$ phases with the in-plane vector along $(100)_{pc}$ and $(010)_{pc}$ with different thickness.

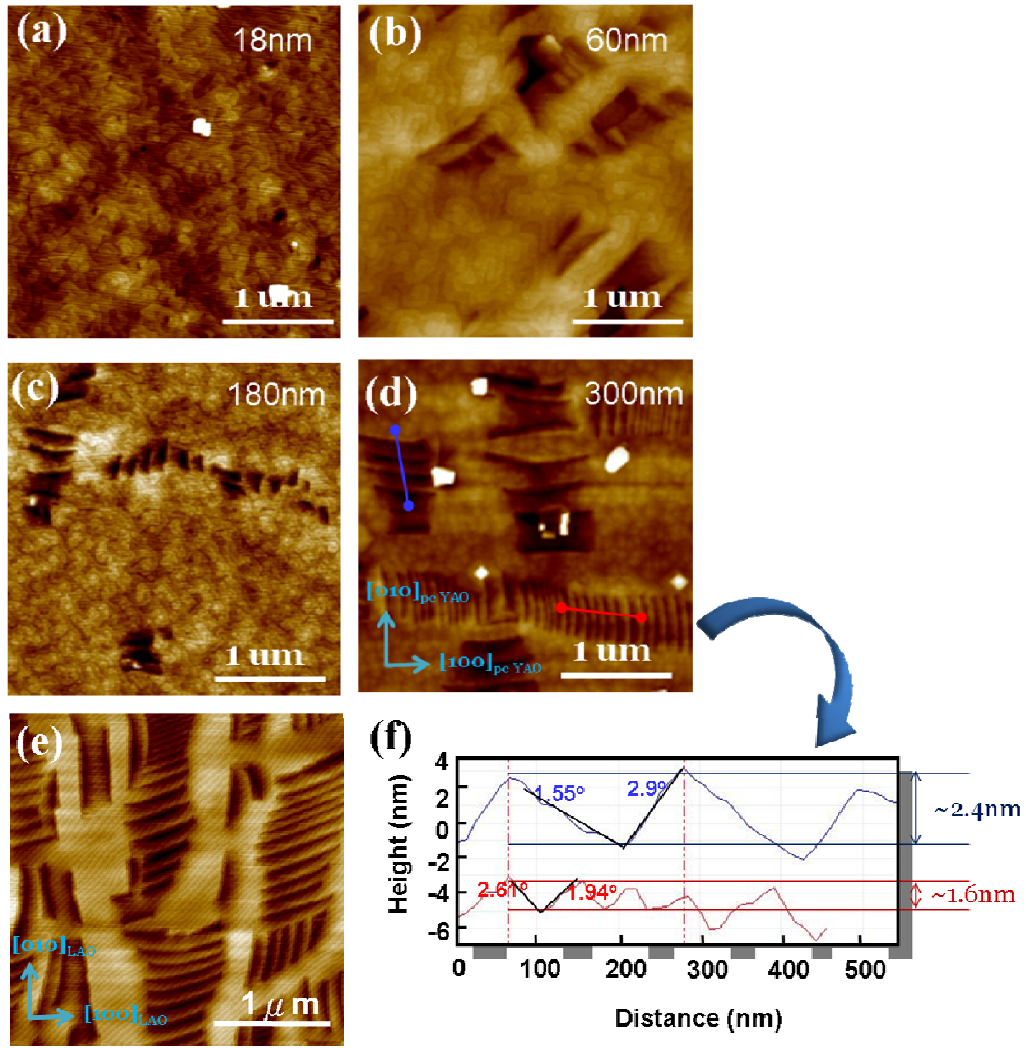

FIG. 5 The topography of (a) 18nm, (b) 60nm, (c) 180nm, and (d) 300nm BFO thin films grown on YAO substrate. (e) 120nm BFO grown on LAO substrate. (f) line-trace along the blue line and red line in (d), which shows different periodic arrangement of strips at $[100]_{pc,YAO}$ and $[010]_{pc,YAO}$. Unlike the uniform arrayed stripes along both LAO [100] and [010] observed in (e), this arrayed stripe anisotropy should results from the lattice difference of a-axis and b-axis of the YAO substrate.

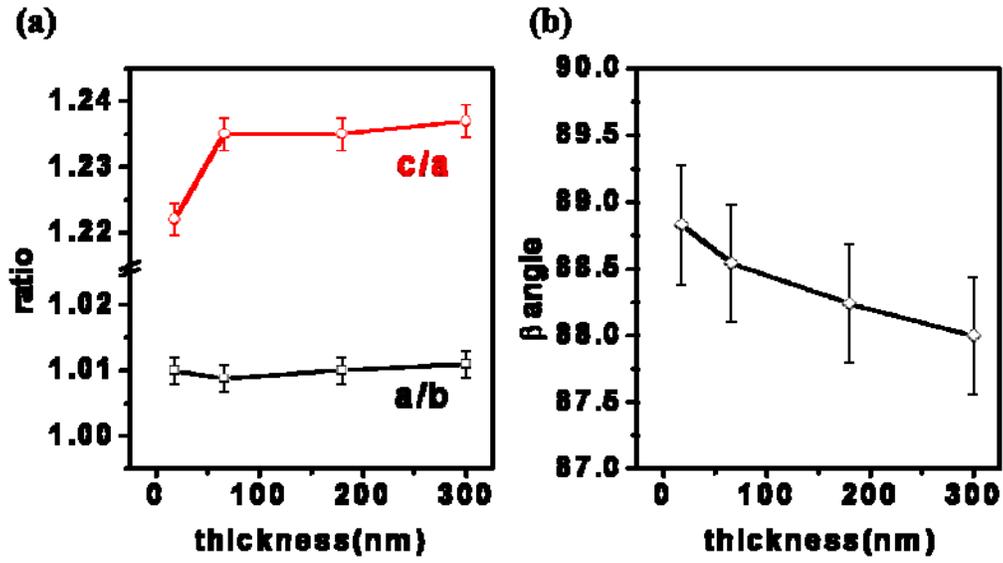

FIG. 6. (a) the ratio of c/a and a/b are calculated from Table. I at different thickness, indicating a stably $M_{II}$ phase can survive at thickness above 300nm. (b) the variation of shear angle of $M_{II}$ phase gradually decreases as thickness increases.